\documentclass[aps,prc,preprint,tightenlines,groupedaddress,nofootinbib,
               showpacs,preprintnumbers,amsmath,amssymb,twocolumn,
               superscriptaddress]{revtex4}
\usepackage{graphicx}
\usepackage{dcolumn}
\usepackage{mathrsfs}
\usepackage{bm}

\begin{document}

\title{Why is Tin so soft?}
\author{J. Piekarewicz}
\affiliation{Department of Physics, Florida State 
             University, Tallahassee, FL 32306}
\date{\today} 

\begin{abstract}
The distribution of isoscalar monopole strength in the neutron-even
${}^{112-124}$Sn-isotopes has been computed using a relativistic
random-phase-approximation approach. The accurately-calibrated model
used here (``FSUGold'') has been successful in reproducing both
ground-state observables as well as collective excitations ---
including the giant monopole resonance (GMR) in ${}^{90}$Zr,
${}^{144}$Sm, and ${}^{208}$Pb. Yet this same model significantly
overestimates the GMR energies in the Sn isotopes. It is argued that
the question of ``Why is Tin so soft?'' becomes an important challenge
to the field and one that should be answered without sacrificing the 
success already achieved by several theoretical models.
\end{abstract}
\maketitle 

The compression modulus of nuclear matter (also known as the nuclear
incompressibility) is a fundamental parameter of the equation of state
that controls small density fluctuations around the saturation
point. While existing ground-state observables have accurately
constrained the binding energy per nucleon ($B/A\!\simeq\!-16$~MeV)
and the baryon density ($\rho\!\simeq\!0.15~{\rm fm}^{-3}$) of
symmetric nuclear matter at saturation, the extraction of the
compression modulus ($K$) requires to probe the response of the
nuclear system to small density fluctuations. It is generally agreed
that the {\sl nuclear compressional modes} --- particularly the
isoscalar giant monopole resonance (GMR) --- provide the optimal route
to the determination of the nuclear
incompressibility~\cite{Blaizot:1980tw}.  Moreover, the field has
attained a level of maturity and sophistication that demands strict
standards in doing so.  It is now demanded that the same microscopic
model that predicts a particular value for the compression modulus of
infinite nuclear matter (an experimentally {\sl inaccessible}
quantity) be able to accurately reproduce the experimental
distribution of monopole strength.

Earlier attempts at extracting the compression modulus of symmetric
nuclear matter relied primarily on the distribution of isoscalar
monopole strength in ${}^{208}$Pb --- a heavy nucleus with a well
developed giant resonance peak~\cite{Youngblood:1977,Youngblood:1981}.
However, as was pointed out recently in
Refs.~\cite{Piekarewicz:2002jd,Piekarewicz:2003br} --- and confirmed
since then by several other
groups~\cite{Vretenar:2003qm,Agrawal:2003xb,Colo:2004mj} --- the GMR
in ${}^{208}$Pb does not provide a clean determination of the
compression modulus of {\sl symmetric} nuclear matter. Rather, it
constraints the nuclear incompressibility of {\sl neutron-rich} matter
at the particular value of the neutron excess found in ${}^{208}$Pb,
namely, $b\!\equiv\!(N\!-\!Z)/A\!=\!0.21$. As such, the GMR in
${}^{208}$Pb is sensitive to the {\sl density dependence of the
symmetry energy}. The symmetry energy represents a penalty levied on
the system as it departs from the symmetric limit of equal number of
neutrons and protons. As the infinite nuclear system becomes neutron
rich, the saturation density moves to lower densities, the binding
energy weakens, and the nuclear incompressibility {\sl
softens}~\cite{Piekarewicz:2007b}. Thus, the compression modulus of a
neutron rich system having the same neutron excess as ${}^{208}$Pb is
{\sl lower} than the compression modulus of symmetric nuclear
matter. We note in passing that the symmetry energy is to an excellent
approximation equal to the difference between the energy of pure
neutron matter (with $b\!\equiv\!1$) and that of symmetric nuclear 
matter (with $b\!\equiv\!0$).

The alluded sensitivity of the distribution of isoscalar monopole
strength to the density dependence of the symmetry energy proved
instrumental in resolving a puzzle involving $K$: how can accurately
calibrated models that reproduce ground state data as well as the
distribution of monopole strength in ${}^{208}$Pb, predict values for
$K$ that differ by as much as 25\%?  This discrepancy is now
attributed to the poorly determined density dependence of the symmetry
energy~\cite{Piekarewicz:2002jd}. Indeed, models that predict a
stiffer symmetry energy (one that increases faster with density)
consistently predict higher compression moduli than those with a
softer symmetry energy. Thus, the success of some models in
reproducing the GMR in ${}^{208}$Pb was accidental, as it resulted
from a combination of both a stiff equation of state for symmetric
nuclear matter and a stiff symmetry energy~\cite{Piekarewicz:2003br}.
Since then, the large differences in the predicted value of $K$ have
been reconciled and a ``consensus'' has been reached that places the
value of the incompressibility coefficient of {\sl symmetric} nuclear
matter at $K\!=\!230\!\pm\!10$~MeV~\cite{Agrawal:2003xb,Colo:2004mj,
Todd-Rutel:2005fa,Garg:2006vc}.

\begin{figure}[ht]
\vspace{0.20in}
\includegraphics[width=3.1in,angle=0]{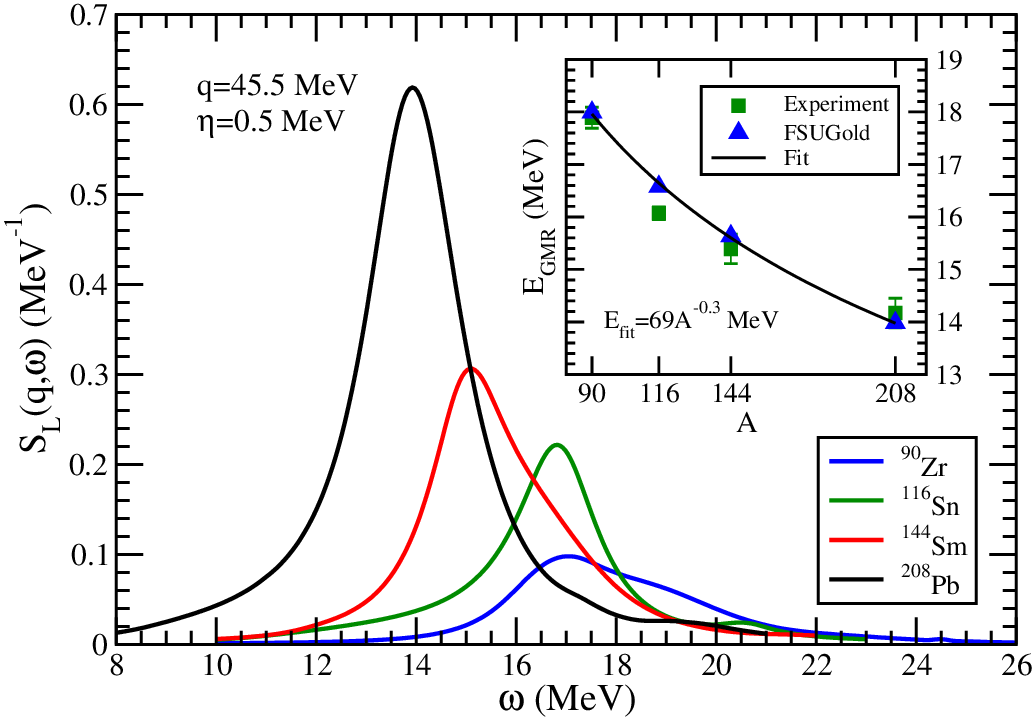}
\caption{(color online) Distribution of isoscalar monopole 
          strength predicted by the FSUGold model of 
          Ref.~\cite{Todd-Rutel:2005fa}. The inset includes 
          a comparison against the experimental centroid 
          energies reported in Ref.~\cite{Youngblood:1999}, 
          with the solid line providing the best fit to the 
          theoretical predictions.}
\label{Fig1}
\end{figure}

An example of how this consensus was reached is depicted in
Fig.~\ref{Fig1} where the distribution of isoscalar monopole strength
in ${}^{90}$Zr, ${}^{116}$Sn, ${}^{144}$Sm, and ${}^{208}$Pb at the
small momentum transfer of $q\!=\!45.5$~MeV (or $q\!=\!0.23~{\rm
fm}^{-1}$) is displayed for the relativistic FSUGold model of
Ref.~\cite{Todd-Rutel:2005fa} --- a model that predicts an
incompressibility coefficient for symmetric nuclear matter of
$K\!=\!230$~MeV.  Note that the distribution of strength was obtained
from a relativistic random-phase-approximation (RPA) approach as
described in detail in Ref.~\cite{Piekarewicz:2001nm}.  Further, the
inset on Fig.~\ref{Fig1} shows a comparison of the theoretical
predictions against the experimental centroid energies reported in
Ref.~\cite{Youngblood:1999}. Finally, the solid line in the inset
provides a fit to the mass dependence of the theoretical predictions
that yields $E_{\rm GMR}(A) \approx [69/A^{0.3}]~{\rm MeV}$.

The isoscalar monopole strength displayed in Fig.~\ref{Fig1} is
extracted from the low momentum transfer behavior of the longitudinal
response defined as follows:
\begin{equation}
  S_{\rm L}({\bf q},\omega) =
  \sum_{n}\Big|\langle\Psi_{n}|\hat{\rho}({\bf q})|
  \Psi_{0}\rangle\Big|^{2}
  \delta(\omega-\omega_{n}) \;.
 \label{SLong}
\end{equation}
Here $\Psi_{0}$ is the exact nuclear ground state, $\Psi_{n}$ is an
excited state with excitation energy $\omega_{n}$, and
$\hat{\rho}({\bf q})$ is the Fourier transform of the isoscalar baryon
density. That is,
\begin{equation}
  \hat{\rho}({\bf q}) = \int d^{3}r \,e^{-i{\bf q}\cdot{\bf r}}
        \bar{\psi}({\bf r})\gamma^{0}\tau_{0}\psi({\bf r})\;,
 \label{RhoQ}
\end{equation}
where $\psi({\bf r})$ is an isodoublet nucleon field, $\gamma^{0}$ is
the timelike (or {\sl zeroth}) component of the Dirac gamma matrices,
and $\tau_{0}\!\equiv\!{\bf 1}$ is the identity matrix in isospin
space. 

The important realization that the distribution of monopole strength
in heavy nuclei is sensitive to the density dependence of the symmetry
energy has motivated a recent experimental study of the GMR along the
isotopic chain in Tin. Indeed, the distribution of isoscalar monopole
strength in the neutron-even ${}^{112-124}$Sn-isotopes has been
measured at the {\sl Research Center for Nuclear Physics}~(RCNP) in
Osaka, Japan~\cite{Garg:2006vc,Garg:2007}. This important experiment
probes the incompressibility of {\sl asymmetric} nuclear matter by
measuring the distribution of isoscalar strength in a chain of
isotopes with a neutron excess ranging from $b\!=\!0.11$ (in
${}^{112}$Sn) to $b\!=\!0.19$ (in ${}^{124}$Sn). The experiment
represents a hadronic compliment to the {\sl purely electroweak}
Parity Radius Experiment (PREX) at the Jefferson Laboratory that aims
to measure the neutron radius of $^{208}$Pb accurately and model
independently via parity-violating electron 
scattering~\cite{Horowitz:1999fk,Michaels:2005}. Such an accurate
determination will have far-reaching implications in areas as diverse
as nuclear structure~\cite{Todd:2003xs}, heavy-ion 
collisions~\cite{Danielewicz:2002pu,Tsang:2004,Chen:2004si,Li:2005sr,
Shetty:2005qp,Horowitz:2006iv}, atomic parity
violation~\cite{Todd:2003xs,Sil:2005tg,Piekarewicz:2006vp} and nuclear
astrophysics~\cite{Horowitz:2000xj,Buras:2003sn,Lattimer:2004pg,
Steiner:2004fi}.

\begin{figure}[ht]
\vspace{0.20in}
\includegraphics[width=3.10in,angle=0]{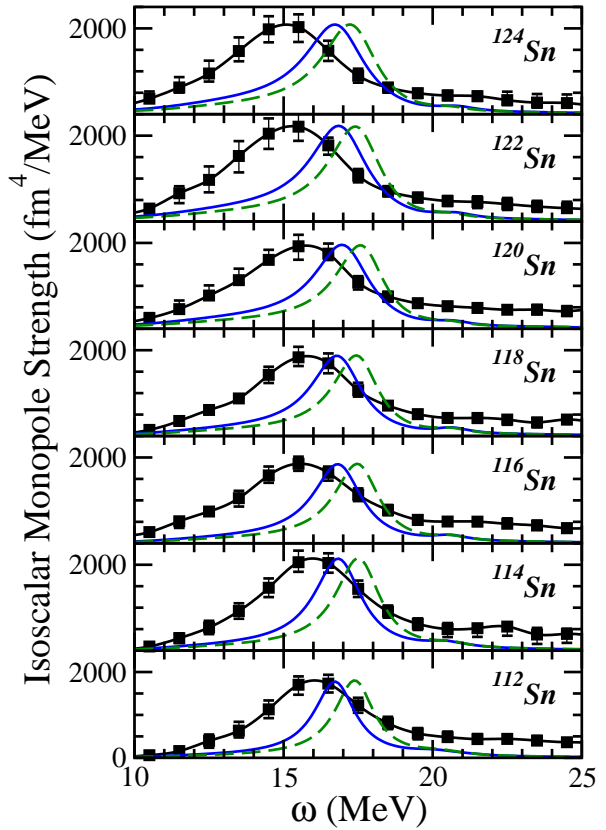}
\caption{(color online) Comparison between the distribution 
          of isoscalar monopole strength in all neutron-even  
          ${}^{112}$Sn-${}^{124}$Sn isotopes extracted from 
          experiment (black solid squares) and the theoretical 
          predictions of the FSUGold (blue solid line) and NL3 
          (green dashed line) models.} 
\label{Fig2}
\end{figure}

\begin{figure}[ht]
\vspace{0.20in}
\includegraphics[width=3.10in,angle=0]{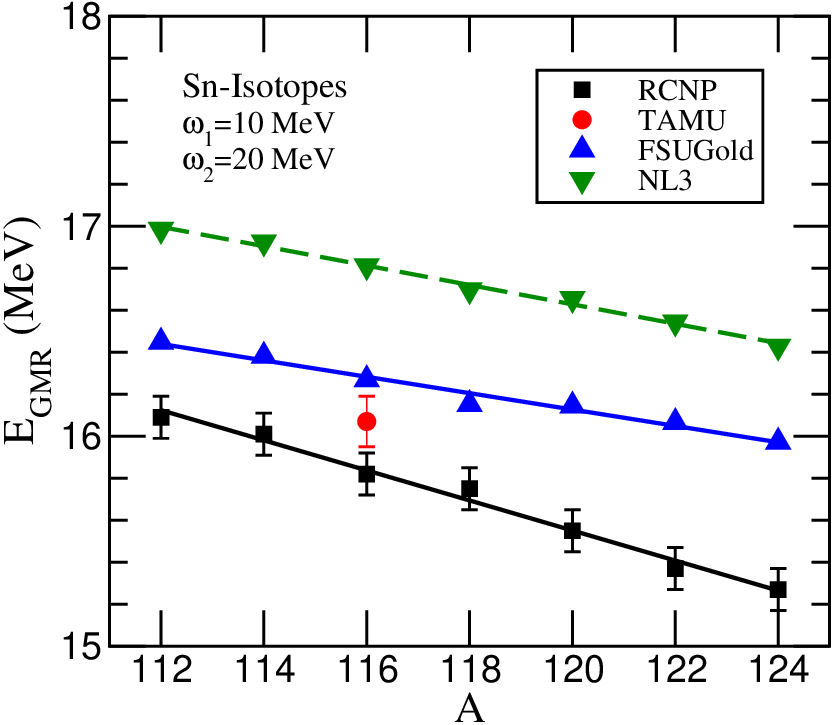}
\caption{(color online) Comparison between the GMR centroid energies
         ($m_{1}/m_{0}$) in all neutron-even ${}^{112}$Sn-${}^{124}$Sn 
         isotopes extracted from experiment (black solid squares) and 
         the theoretical predictions of the FSUGold (blue up-triangles) 
         and NL3 (green down-triangles) models. Also shown (filled
         red circle) is the result of Ref.~\cite{Youngblood:1999} 
         for the case of ${}^{116}$Sn.} 
\label{Fig3}
\end{figure}

In Fig.~\ref{Fig2} the experimental distribution of isoscalar monopole
strength measured at the RCNP~\cite{Garg:2006vc,Garg:2007} is compared
against the predictions of the highly successful
NL3~\cite{Lalazissis:1996rd,Lalazissis:1999} and
FSUGold~\cite{Todd-Rutel:2005fa} models. As one is only interested in
comparing the shape of the distribution and a particular ratio of its
moments, the maximum of the theoretical curves --- computed from the
longitudinal response as described in the text --- has been normalized
to the experimental data.  The $A$-dependence of the corresponding
centroid energies is also displayed in Fig.~\ref{Fig3} and compiled in
Table~\ref{Table1}. Note that the centroid energy is computed from the
ratio of the $m_{1}$ moment to that of the $m_{0}$ moment. That is,
\begin{equation}
  E_{\rm GMR} \equiv \frac{m_{1}}{m_{0}} =
  \frac{\int_{\omega_{1}}^{\omega_{2}}\omega S_{\rm L}(q_{0},\omega)d\omega}
       {\int_{\omega_{1}}^{\omega_{2}}S_{\rm L}(q_{0},\omega)d\omega} \;,
 \label{Centroid}
\end{equation}
where, consistent with the experimental analysis~\cite{Garg:2006vc,
Garg:2007}, the limits of integration have been chosen to be
$\omega_{1}\!=\!10$~MeV and $\omega_{2}\!=\!20$~MeV. Further, to mimic 
the forward-angle experiment, the longitudinal response was evaluated
at the ``small'' momentum transfer of $q_{0}\!=\!0.23~{\rm fm}^{-1}$.

\begin{table}
\begin{tabular}{|c|c|c|c|}
 \hline
  Nucleus & NL3 & FSUGold & Experiment \\
 \hline
 \hline
  ${}^{112}$Sn & $16.98$ & $16.45$ & $16.2\!\pm\!0.1$ \\
  ${}^{114}$Sn & $16.92$ & $16.38$ & $16.1\!\pm\!0.1$ \\
  ${}^{116}$Sn & $16.81$ & $16.27$ & $15.8\!\pm\!0.1$ \\
  ${}^{118}$Sn & $16.70$ & $16.15$ & $15.8\!\pm\!0.1$ \\
  ${}^{120}$Sn & $16.66$ & $16.14$ & $15.7\!\pm\!0.1$ \\
  ${}^{122}$Sn & $16.54$ & $16.07$ & $15.4\!\pm\!0.1$ \\
  ${}^{124}$Sn & $16.43$ & $15.97$ & $15.3\!\pm\!0.1$ \\
\hline
\end{tabular}
\caption{Giant Monopole Resonance centroid energies (in MeV)
         computed from the ratio of moments ($m_{1}/m_{0}$) 
         as described in the text. All moments were obtained
         from integrating the distribution of strength over
         the $10\le\omega\le20$ MeV interval.} 
\label{Table1}
\end{table}

A subtle telltale problem with Tin barely discernible in the inset on
Fig.~\ref{Fig1}, becomes magnified in Fig.~\ref{Fig2} as one compares
the experimentally extracted distribution of monopole strength against
the theoretical predictions.  While mean-field plus RPA calculations
are typically unable to describe the experimental width --- which is
in general composed of both an escape (particle-hole) and a spreading
(multiparticle-multihole) width --- such is not the case for the
description of the centroid energies. Indeed, accurately calibrated
models, both non-relativistic~\cite{Colo:2004mj} and relativistic (see
Fig.~\ref{Fig1}), provide an adequate description of the GMR centroid
energies in both ${}^{90}$Zr (with $b\!=\!0.11$) and $^{208}$Pb (with
$b\!=\!0.21$) --- nuclei with a neutron excess similar to those at the
two extremes of the isotopic chain considered here. Why is then that
both non-relativistic~\cite{Garg:2006vc,Garg:2007,Colo:2007} and
relativistic models consistently overestimate the centroid energies in
the Sn-isotopes? Or more colloquially, {\sl why is Tin so soft?} And
why is that the discrepancy between theory and experiment continues to
grow as the neutron excess increases? A stiff symmetry energy leads to
a rapid softening of the nuclear
incompressibility~\cite{Piekarewicz:2007b}. This is the main reason
behind the slightly larger (negative) slope displayed by NL3 relative
to FSUGold in Fig.~\ref{Fig3}. The even larger (by more than 50\%)
slope displayed by the experimental data is unlikely to be solely
related to the stiffness of the symmetry energy, as NL3 already
predicts a neutron skin thickness in $^{208}$Pb that appears overly
large~\cite{Todd-Rutel:2005fa}.

So why is Tin so soft and why does it become even softer with an
increase in the neutron excess?  Could there be a systematic error in
the experimental extraction? While possible, this is unlikely as an
earlier independent measurement on $^{116}$Sn~\cite{Youngblood:1999}
appears to confirm the present (RCNP) result (see
Fig.~\ref{Fig3}). Could the GMR in Tin probe physics that has not been
already constrained by nuclear observables?  This also appears
unlikely as existing density functionals are successful at describing
a host of ground-state observables as well as collective excitations
--- including the GMR in ${}^{90}$Zr, ${}^{144}$Sm, and ${}^{208}$Pb
(see Fig.~\ref{Fig1} and Ref.~\cite{Todd-Rutel:2005fa}). Could Tin be
sensitive to pairing correlations and more complicated
multiparticle-multihole excitations? The answer appears to be
negative~\cite{Colo:2007}, but even if it would be positive, why
should Tin be sensitive to these effects but not Zr, Sm, and Pb?
Clearly, the distribution of isoscalar monopole strength in the
Sn-isotopes poses a serious theoretical challenge, perhaps suitable
for the new Universal Nuclear Energy Density Functional (UNEDF)
initiative. Whatever the theoretical approach, however, one must
remember that the challenge is not solely to describe the distribution
of monopole strength along the isotopic chain in Tin, but rather to
do so without sacrificing the enormous success already achieved in
reproducing a host of ground-state properties and collective modes.

\begin{acknowledgments}
The author is grateful to Professors G. Col\`o and U. Garg for many
fruitful discussions. The author also wishes to thank Prof. Garg and
his collaborators for sharing the experimental data prior to
publication. This work was supported in part by DOE grant
DE-FD05-92ER40750.
\end{acknowledgments}

\vfill
\bibliography{ReferencesJP}

\end{document}